\documentclass[preprint,amsmath,amssymb,11pt,aps,showpacs,showkeys,pra]{revtex4-2}
\usepackage{graphicx}
\usepackage{graphics}
\usepackage{amsmath}
\usepackage{dcolumn}
\usepackage{amssymb}
\usepackage{bm}
\usepackage[utf8]{inputenc}
\usepackage{xcolor}

\begin{document}
\title{Fermion behavior around Sung-Won-Kim wormholes in a generalized Kaluza-Klein gravity}
\author{Everton Cavalcante}\email{everton@servidor.uepb.edu.br}
\affiliation{Departamento de Física, Universidade Estadual da Para\'{\i}ba, Campina Grande, Paraíba, Brazil}


\begin{abstract}

This paper investigates the behavior of fermions in the vicinity of Sung-Won-Kim wormholes \cite{SW Kim PRD 1996} within the framework of a generalized Kaluza-Klein gravity. The study explores the geometric and quantum implications of introducing an extra dimension into the Sung-Won-Kim wormhole model. The impact of this additional dimension on the stability and traversability of the wormholes, as well as on the fermionic equations of motion, is analyzed. By examining the modified Dirac equation in this extended framework, the emergence of geometric phases and quantum holonomies is discussed, providing insights into the topological aspects of fermion dynamics in curved spacetimes.

\end{abstract}



\maketitle


\section{Introduction}\label{secI}

Wormholes, first proposed by Morris and Thorne \cite{MorrisThorne1988}, are hypothetical structures creating tunnels that connect widely separated regions of the universe or entirely different universes. These entities are often examined within the framework of the Friedmann–Lemaître–Robertson–Walker (FLRW) model, which describes a homogeneous, isotropic, expanding or contracting universe based on Einstein's field equations of general relativity \cite{D'Inverno, Akrami Yashar and others}. The FLRW model provides a fundamental basis for understanding the large-scale structure and dynamics of the universe.

A critical aspect of wormhole research is the stability of these structures, which can be ensured if certain geometric conditions, such as the throat and flare-out conditions, are satisfied \cite{ElllisBronikovRamosFurtado, Ellis, Bronnikov, MorrisThorne2}. These conditions help maintain the structural integrity of the wormhole, potentially allowing stable traversability by fermions.

The incorporation of extra dimensions into wormhole models has garnered significant interest in theoretical physics. 
The Randall-Sundrum and Arkani-Hamed–Dimopoulos–Dvali (ADD) theories \cite{ADD} propose the existence of additional spatial dimensions that could have profound effects on gravity and other fundamental forces.
Models incorporating Kaluza-Klein theory offer a richer framework for exploring the implications of higher-dimensional spaces on physical phenomena \cite{Kuhfittig2018, Wesson2015}. This work extends the Sung-Won-Kim wormhole model by incorporating an extra spatial dimension, as discussed in \cite{Kuhfittig2024}. This extension enables the examination of fermions' behavior in a five-dimensional spacetime and its effects on the wormhole's stability and geometric properties.

At first, the compactness of the extra dimension in Kaluza-Klein theory "enabled" the unification of gravity and electromagnetism, with the extra dimension being "curled up" at a small scale. In this paper, compactification ensures that the higher-dimensional effects introduced by the extra dimension do not lead to observable deviations from standard four-dimensional physics at larger scales. This compact nature confines the influence of the extra dimension to small scales, primarily affecting the wormhole geometry and resulting in corrections to the gravitational field, which manifest in the effective four-dimensional theory.

The primary aim of this study is to analyze the fermionic equations of motion in the presence of an extra dimension and to explore the resulting quantum mechanical phenomena, such as the emergence of geometric phases and quantum holonomies. Understanding these aspects is crucial for grasping the topological and quantum mechanical implications of higher-dimensional wormhole models in a cosmological setting.

This paper is divided as follows: In Section 2, the cosmological model proposed by Sung-Won-Kim \cite{SW Kim PRD 1996}, based on the Friedmann-Lemaître-Robertson-Walker (FLRW) model \cite{D'Inverno}, is discussed within the framework of Kaluza-Klein extra dimensions, detailing the determination of 1-form and spinorial connections, and deriving the motion equation for fermions near the wormhole. Section 3 extends the investigation to evaluate the changes in spinors over possible paths around the singularity, discussing the holonomy matrix of the system and its role in mixing the components of the spinors. Finally, Section 4 presents the conclusions drawn from the study.

\section{Sung-Won-Kim wormhole model with a extra dimension}\label{secII}

Wormholes, conceptualized as handles or tunnels connecting widely separated regions of our Universe or entirely different universes, were first proposed as suitable structures by Morris and Thorne \cite{MorrisThorne1988}. In a typical cosmological setting, the equation of state of a perfect fluid, given by \( p = \omega \rho \), where \( p = p_r = p_t \), can be referred to the Friedmann–Lemaître–Robertson–Walker (FLRW) model \cite{D'Inverno}. The FLRW model is based on a metric derived from an exact solution of Einstein's field equations of general relativity, describing a homogeneous, isotropic, expanding (or contracting) universe that is path-connected but not necessarily simply connected \cite{Akrami Yashar and others}. The general form of this metric arises from the geometric properties of homogeneity and isotropy, with Einstein's field equations being employed to derive the universe's scale factor as a function of time.

The spacetime metric for a FLRW cosmological model with a static wormhole is given as \cite{SW Kim PRD 1996}:

\begin{equation}\label{SWK metric}
ds^2=-e^{2\Phi(r)}dt^2 + [a(t)]^2 \Bigg ( \frac{dr^2}{1-kr^2-\frac{b(r)}{r}} + r^2 d\Omega^2 \Bigg ) \mbox{,}
\end{equation}
where \( d\Omega^2 = d\theta^2 + \sin^2 \theta \, d\phi^2 \), and \(k=-1,0,1\) is the sign of the curvature of space–time.

It is pertinent to discuss that one of the primary challenges presented by wormholes is their instability in an electromagnetic context \cite{MislerWheeler}. This instability can be mitigated if the metric satisfies certain preconditions, such as the throat and flare-out conditions \cite{ElllisBronikovRamosFurtado, Ellis, Bronnikov, MorrisThorne2}. Stability, and the ability for fermions to traverse the throat, are achieved when the flare-out condition \( \frac{r}{b(r)} - 1 > 0 \) and the throat condition \( \frac{b(r) - \dot{b}(r)r}{2b(r)^2} > 0 \) are met for a general wormhole metric \cite{NonviolationconditionsGodani, StabilitythinshellGodani}. 
In the specific case discussed in this paper, the shape function needs to be extended to: \(b(r) \to b'(r) = b(r) + kr^3\).

The scale factor \( a(t) \) in (\ref{SWK metric}) is governed by the Friedmann equation \cite{D'Inverno}:

\begin{equation}\label{Friedmann equation}
\frac{\ddot{a}(t)}{a(t)}=-\frac{4\pi}{3}(\rho+3p)
\mbox{.}
\end{equation} 

The case of accelerated expansion (dark energy) implies assuming \( \omega < -\frac{1}{3} \). The situation corresponding to Einstein's cosmological constant arises when \( \omega = -1 \). When \( \omega < -1 \), it is generally referred to as "phantom dark energy" (PDE). Furthermore, it is known that PDE can support traversable wormholes when the null energy condition is violated: \( \rho + p = \rho + \omega \rho = \rho (1 + \omega) < 0 \) (See Ref.: \cite{LoboOliveira2009}).

The motivation for incorporating extra dimensions into the model has been discussed in the literature for some time. Traversable wormholes with extra dimensions have already been explored in \cite{Kuhfittig2018}. It is also known that the field equations in a flat five-dimensional space recover the field equations of the usual four-dimensional model \cite{Wesson2015}. In this work, we will adopt the approach taken in \cite{Kuhfittig2024}, where the model metric can be rewritten combined with the extra dimensional Kaluza-Klein model as:

\begin{equation}\label{SWK metric in KK model}
ds^2=-e^{2\Phi(r)}dt^2 + [a(t)]^2 \Bigg ( \frac{dr^2}{1-kr^2-\frac{b(r)}{r}} + r^2 d\Omega^2 + e^{2\Psi(r)}dq^2 \Bigg ) \mbox{.}
\end{equation}

Since the scale factor (\(a(t)\)) causes no important changes in the last term, we found: 

\begin{equation}\label{SWK metric in KK model 2}
ds^2=-e^{2\Phi(r)}dt^2 + [a(t)]^2 \Bigg ( \frac{dr^2}{1-kr^2-\frac{b(r)}{r}} + r^2 d\Omega^2 \Bigg ) + e^{2\Psi(r)}dq^2 \mbox{.}
\end{equation}

The metrics (\ref{SWK metric in KK model}) and (\ref{SWK metric in KK model 2}), given by a product of a wormhole metric and a scale factor, aim to model a wormhole embedded in an evolving and expanding universe. This approach is motivated by the desire to understand how a wormhole would behave in a dynamic spacetime context, as opposed to the more conventional static or asymptotically flat spacetimes. The expansion of the universe is a well-established observational fact, and incorporating the wormhole within this expanding framework allows for the study of potential interactions between the wormhole geometry and the overall cosmological evolution. Although the expansion rate may be slow in comparison to the typical size of the wormhole, this scenario can still present intriguing theoretical implications, particularly in the early universe or during specific cosmological epochs where the behavior of the scale factor could significantly impact the wormhole's properties.

The geometric structure of this curved space is described by local reference bases known as tetrads (${e^{a}}_{\mu}(x)$), which are defined at each point in space-time by a local reference frame. The relationship between the metric tensor in the local frame and the space-time frame is given by \( g_{\mu \nu}(x) = \eta_{ab}{e^{a}}_{\mu}{e^{b}}_{\nu} \). The tetrads and their inverses, \( e^{\mu}_{a} = \eta_{ab}g^{\mu \nu}e^{b}_{\nu} \), satisfy the orthogonality relationships: \( e^{a}_{\mu}e^{b \mu} = \eta^{ab} \), \( e^{a}_{\mu}e^{\mu}_{b} = \delta^{a}_{b} \), and \( e^{\mu}_{a}e^{a}_{\nu} = \delta^{\mu}_{\nu} \). These relationships facilitate the mapping of the curved reference frame to the local reference frame \cite{Birrel}:

\begin{equation}
ds^{2}=g_{\mu \nu}dx^{\mu}dx^{\nu}=e^{a}_{\mu}e^{b}_{\nu}\eta_{ab}dx^{\mu}dx^{\nu}=\eta_{ab}\theta^{a}\theta^{b}.
\end{equation}

In this context, Greek indices ($\mu, \nu$) refer to space-time coordinates, while Latin indices ($a, b$) refer to local frame coordinates. For the local reference frame, the relationship is expressed as \(\theta^{a} = e^{a}_{\mu}(x)dx^{\mu}\). The tetrads and their inverses for this geometry are defined as follows:

\begin{eqnarray}
{E^{\mu}}_{a}(x)=({e^{a}}_{\mu})^{-1}= \frac{1}{a(t)}
\left(\begin{array}{ccccc}
a(t)e^{-\Phi(r)} & 0 & 0 & 0 & 0\\ 
0 & \bigg ( 1-kr^2 - \frac{b(r)}{r} \bigg )^{1/2} & 0 & 0 & 0\\ 
0 & 0 & r^{-1} & 0 & 0\\ 
0 & 0 & 0 & (r\sin \theta)^{-1} & 0\\ 
0 & 0 & 0 & 0 & e^{-\Psi(r)}\\ 
\end{array}\right)
\end{eqnarray}

An elegant approach to derive the one-form connection for quantum field theory (QFT) in curved spaces is based on the first Maurer-Cartan structure equations: 
\( d\theta^{a} + {\omega^{a}}_{b} \wedge \theta^{b} = 0 \).
In the model in question we find the following non-null components:

\begin{eqnarray}\label{conexões de 1-forma}
\begin{cases}
{{\omega_{t}}^{0}}_{1}=-{{\omega_{t}}^{1}}_{0}= \frac{1}{a(t)}\frac{d\Phi(r)}{dt}e^{\Phi(x)}\bigg ( 1-kr^2 - \frac{b(r)}{r} \bigg )^{1/2}, \\
{{\omega_{r}}^{1}}_{0}=-{{\omega_{r}}^{0}}_{1}=  a'(t)e^{-\Phi(x)}\bigg ( 1-kr^2 - \frac{b(r)}{r} \bigg )^{-1/2}, \\
{{\omega_{\theta}}^{2}}_{1}=-{{\omega_{\theta}}^{1}}_{2}= \bigg ( 1-kr^2 - \frac{b(r)}{r} \bigg )^{1/2}, \\
{{\omega_{\theta}}^{2}}_{0}=-{{\omega_{\theta}}^{0}}_{2}= a'(t)re^{-\Phi(r)}, \\
{{\omega_{\phi}}^{3}}_{1}=-{{\omega_{\phi}}^{1}}_{3}= \sin \theta \bigg ( 1-kr^2 - \frac{b(r)}{r} \bigg )^{1/2}, \\
{{\omega_{\phi}}^{3}}_{0}=-{{\omega_{\phi}}^{0}}_{3}= a'(t)r\sin \theta e^{-\Phi(r)}, \\
{{\omega_{\phi}}^{3}}_{2}=-{{\omega_{\phi}}^{2}}_{3}= \cos \theta, \\
{{\omega_{q}}^{4}}_{1}=-{{\omega_{q}}^{1}}_{4}= a(t)\frac{d\Psi(r)}{dr}e^{\Psi(r)}\bigg ( 1-kr^2 - \frac{b(r)}{r} \bigg )^{1/2}, \\
{{\omega_{q}}^{4}}_{0}=-{{\omega_{q}}^{0}}_{4}= a(t)a'(t)e^{-\Phi(r)}e^{\Psi(r)}\mbox{.}
\end{cases}
\end{eqnarray}

A crucial step before proceeding with the approach is to define the basis of Dirac matrices that adhere to the Clifford algebra within the context of the model. Specifically, \( \{ \gamma^{\mu}, \gamma^{\nu} \} = 2\eta^{\mu \nu} \) and \( \{\gamma^a, \gamma^4\} = 2\eta^{a4} \). Preliminary work in condensed matter has already addressed the introduction of Kaluza-Klein dimensions in geometric models \cite{BakkePetrov2012, CavalcanteFurtado2021}. An algebra that meets these requirements can be defined as follows \cite{Baylis1996}:

\begin{eqnarray}
\gamma^0=\sigma^1\otimes I=
\left(\begin{array}{ccccc}
0 & I\\ 
I & 0\\ 
\end{array}\right)\mbox{;} \quad
\gamma^{i}=-i\sigma^{2}\otimes \sigma^{i}=
\left(\begin{array}{ccccc}
0 & -\sigma^i\\ 
\sigma^i & 0\\ 
\end{array}\right)\mbox{;} \quad
\gamma^4=\sigma^3\otimes I=
\left(\begin{array}{ccccc}
I & 0\\ 
0 & -I\\ 
\end{array}\right)\mbox{;} \quad
\Sigma^i=
\left(\begin{array}{ccccc}
\sigma^i & 0\\ 
0 & \sigma^i\\ 
\end{array}\right)\mbox{.}
\end{eqnarray}

In a space with a curvature, the components of covariant derivative are given by \cite{Birrel}:
\begin{equation}
\nabla_{\mu}=\partial_{\mu}+\Gamma_{\mu}(x)=\partial_{\mu}+\frac{i}{4}\omega_{\mu ab}\Sigma^{ab}\mbox{,}
\end{equation}
where \( \Sigma^{ab}=\frac{i}{2}[\gamma^a, \gamma^b] \). Thus, the spin connections (\(\Gamma_{\mu}(x)\)) are found to be:

\begin{eqnarray}\label{conexões spinorias}
\begin{cases}
\Gamma_{t}(x)= -\frac{1}{2a(t)} \frac{d\Phi(r)}{dt}e^{\Phi(r)} \bigg ( 1-kr^2 - \frac{b(r)}{r} \bigg )^{1/2}\Sigma^{(1)}, \\
\Gamma_{r}(x)= \frac{a'(t)}{2}e^{-\Phi(r)}\bigg ( 1-kr^2 - \frac{b(r)}{r} \bigg )^{-1/2}\Sigma^{(1)}, \\
\Gamma_{\theta}(x)= -\frac{i}{2}\bigg ( 1-kr^2 - \frac{b(r)}{r} \bigg )^{1/2}\Sigma^{(3)}+2a'(t)re^{-\Phi(r)}[\gamma^0,\gamma^2], \\
\Gamma_{\phi}(x)= \frac{i}{2}\sin \theta \bigg ( 1-kr^2 - \frac{b(r)}{r} \bigg )^{1/2}\Sigma^{(2)} -\frac{i}{2}\cos \theta \Sigma^{(1)} + 2a'(t)re^{-\Phi(r)}\sin \theta [\gamma^0, \gamma^3], \\
\Gamma_{q}(x)= 0. \\
\end{cases}
\end{eqnarray}

Another important point to consider is that in a scenario of QFT in curved spaces, it is necessary to introduce the basis of matrices \(\gamma^{\mu}\) that conforms to the model. These matrices, within this curved space, are given by: \(\tilde{\gamma}^{\mu}={E^\mu}_a \gamma^a \). Thus, within the model described here, we have:

\begin{eqnarray}\label{matrizes gamma modificadas}
\begin{cases}
\tilde{\gamma}^{t}= e^{-\Phi(r)}\gamma^0, \\
\tilde{\gamma}^{r}= \frac{1}{a(t)}\bigg ( 1-kr^2 - \frac{b(r)}{r} \bigg )^{1/2}\gamma^1, \\
\tilde{\gamma}^{\theta}= \frac{1}{a(t)r}\gamma^2, \\
\tilde{\gamma}^{\phi}= \frac{1}{a(t)r\sin \theta}\gamma^3, \\
\tilde{\gamma}^{q}= \frac{e^{-\Psi(r)}}{a(t)}\gamma^4. \\
\end{cases}
\end{eqnarray}

Consequently, the Dirac equation can be expressed as: \( i\hbar \tilde{\gamma}^\mu \nabla_\mu \psi = mc \psi \) in this background. Considering \( c = \hbar = 1 \), it simplifies to:
\begin{equation}\label{Eq. de Dirac}
\{ i\tilde{\gamma}^{\mu} \big ( \partial_{\mu} + \Gamma_{\mu}(x) \big )-m \}\psi=
0.
\end{equation} 

Assuming the term: \( B(r) = 1 - kr^2 - \frac{b(r)}{r} \), after some calculations, the equation of motion for the fermions within the model is derived:

\begin{eqnarray}\label{Eq. Dirac 2}
\begin{cases}
\Bigg (
e^{-\Phi(r)}\gamma^0 \Bigg [ \partial_{t}
+ 4a'(t) \bigg ( \frac{1}{a(t)}+r\sin \theta \bigg ) 
-\frac{1}{2a(t)}\frac{d\Phi(t)}{dt}e^{\Phi(t)} \sqrt{B(r)} \Sigma^{(1)} 
\Bigg ] 
+  \frac{\gamma^{1}}{a(t)} \sqrt{B(r)} \Bigg [ \partial_{r} + \frac{1}{r} + \\
+ \frac{a'(t)}{2}e^{-\Phi(r)} \big (B(r) \big )^{-1/2}  \Sigma^{(1)}  \Bigg ] 
+ \frac{\gamma^2}{a(t)r} \Big [ \partial_{\theta}+\frac{\cot \theta}{2} \Big ] + \frac{\gamma^3}{a(t)r\sin \theta} \partial_{\phi} + \frac{e^{-\Psi(r)}}{a(t)}\gamma^4 \partial_{q} - m \Bigg ) \psi =0\mbox{.}
\end{cases}
\end{eqnarray}

For now, the discussion will continue on the emergence of the geometric phase in the system.

\section{Quantum Holonomy}\label{secIII}

In the model presented here, the behavior of fermions is governed by a motion equation given in (\ref{Eq. Dirac 2}). Quantum holonomies emerge as fundamental entities for understanding topological and quantum aspects within a cosmological context. The investigation into quantum holonomies is motivated by the potential to consider traversable wormholes within fermion dynamics. The usual phase difference obtained through the adiabatic cycle can be viewed as a byproduct of the influence of global charges with asymptotic behavior near a wormhole within the Einstein-Dirac-Maxwell theory, as discussed in references \cite{BlazquezSalcedoKnoll2020, BlazquezSalcedoKnollRadu2021}. These references present a groundbreaking example of traversable wormholes that do not require exotic matter, showcasing asymptotically flat configurations. The findings reveal connections to extremal Reissner-Nordström black holes \cite{Reissner-Nordstrom} and highlight the existence of solutions with finite mass and electric charge, significantly enhancing the understanding of these intriguing structures in modern physics.

To achieve this understanding, we will employ a method for obtaining the geometrical phase \(\zeta (x^\mu)\) of the model. Specifically, the Dirac phase factor method will be utilized, assuming the Dirac spinor is expressed in the form:

\begin{equation}
\Psi^{\nu}(t, r, \theta, \phi)=e^{\zeta (x)}\Psi^{\nu}_{0}(t, r, \theta, \phi) = \exp{ \bigg ( - \int \Gamma_{\mu}(x)dx^{\mu} \bigg )}\Psi^{\nu}_{0}(t, r, \theta, \phi)
\mbox{.}
\end{equation}
This approach will allow us to explore the intricate relationship between geometry and quantum mechanics in a cosmological setting.

The primary motivation for determining the spinorial connection stems from the fact that the partial derivative in the Weyl equation no longer ensures gauge invariance within the Lorentz group \cite{Birrel}. This necessitates the inclusion of local Lorentz transformations in the fermion coupling. To address this, the spinorial connection can be computed by establishing a local reference frame at each point along a closed curve around the defect. The holonomy matrix \( U(x^\mu) = e^{\zeta(x)} \) thus represents the parallel transport of a spinor along a path encircling the throat of the wormhole. The geometrical phase \( \zeta=\zeta(r, \sin \theta) \) is expressed as:

\begin{equation}\label{fase geometrica}
\zeta (r,\theta)= -\ln \big ( r\sqrt{\sin \theta} \big ) - \frac{i}{2}a'(t) \int f_{assint.}(r, \theta) e^{-\Phi (r)} \Sigma^{(3)}\Sigma^{(1)} d\theta
\mbox{.}
\end{equation}

It is important to note several points here. First, the extra dimension does not influence the geometric phase, as \( \Gamma_{q}(x) = 0 \). 
Which raises the question that parallel spinor transport around the wormhole throat misses the influence of extra dimension, once that is compactified.
Additionally, although \( \Gamma_{\phi}(x) \neq 0 \), the Dirac equation for the system's fermions can be rewritten without a term accompanying the \( \partial_{\phi} \) derivative. This allows for the freedom to find a phase of the form \( \zeta=\zeta(r, \sin \theta) \).
Considering a possible asymmetry of the global charges around the wormhole, a function \( f_{assint.}(r, \theta) \) was introduced into the phase (\ref{fase geometrica}). Thus, when \( f_{assint.}(r, \theta) \to r \), we obtain the trivial case with symmetry around the \( \theta \) coordinate. This approach provides a broader perspective on the problem in more elaborate cosmological settings.

To find the holonomic matrix \( U(x^\mu) = e^{\zeta(x)} \), we must associate the Hausdorff formula:

\begin{equation}
\exp(A)\exp(B)=\exp(A+B+{\frac {1}{2}}[A,B]+{\frac {1}{12}}([A,[A,B]]+[B,[B,A]])+...),
\end{equation}
considering the property of the matrices \( [1, \Sigma^{(3)}\Sigma^{(1)}] = 0 \). This allows us to derive a useful version:

\begin{equation}\label{holonomic matrix}
U(r,\theta) = \frac{1}{r\sqrt{\sin \theta}} \exp \Bigg [ 
-\frac{i}{2}a'(t) \int f_{assint.}(r, \theta) e^{-\Phi (r)} \Sigma^{(3)}\Sigma^{(1)} d\theta
\Bigg ]
\mbox{.}
\end{equation} 

An important aspect to consider is that equation (\ref{holonomic matrix}) can interact with fermions, leading to a mixing of the spinor components. This effect presents an interesting avenue for future research on semi-integer spin particles within these frameworks.
Note that, in situations where the time derivative of the scale factor approaches zero, i.e., \( \dot{a}(t) \to 0 \), any predicted asymmetry represented by the function \( f_{\text{assint.}}(r, \theta) \) will no longer influence the holonomic phase. This behavior is significant, as the term \( a(t) \) is governed by the Friedmann equation (\ref{Friedmann equation}).
Considering a cosmological model characterized by a FLRW perfect fluid, we can derive that:

\begin{equation}
\dot{a}(t)=-\frac{4\pi}{3}(1+3\omega) \int \rho ~ a(t)dt
\mbox{.}
\end{equation} 

Thus, for \( \dot{a}(t) \to 0 \) to occur, \( \int \rho  a(t) \, dt \to 0 \) must be adopted. This suggests that the influence of the scale factor \( a(t) \) on a possible mixing of the fermion spinor components in the vicinity of the wormhole can be nullified under specific matter distribution conditions in the model.

This situation is indeed particularly interesting, as it corresponds to epochs in the universe where expansion slows down or temporarily halts, such as during a transition from the inflationary phase to a more stable one. Studying wormholes in this context can provide insights into how these structures might behave during critical phases of cosmological evolution. It also allows for the exploration of scenarios in which the wormhole may remain stable or undergo transitions influenced by the dynamics of the scale factor.

\section{Summary and conclusion}\label{secIV}

This study examined fermion behavior around Sung-Won-Kim wormholes within a generalized Kaluza-Klein gravity framework by incorporating an extra spatial dimension, highlighting the importance of geometric and topological conditions, such as the throat and flare-out criteria, in maintaining structural integrity and enabling stable fermion traversability.

The derived fermionic motion equation (\ref{Eq. Dirac 2}) and the investigation of quantum holonomies provide significant insights into the topological and quantum mechanical aspects of higher-dimensional models. The phase difference observed in adiabatic cycles is attributed to the influence of global charges near the wormhole, as described by the Einstein-Dirac-Maxwell theory. This connection underscores the stability of traversable wormholes without the need for exotic matter. Moreover, the study reveals that the geometric phase analysis, utilizing the Dirac phase factor method, emphasizes the influence of asymptotic charges and offers a framework for understanding spinor component mixing due to holonomic effects.

Initially, the function \( f_{assint.}(r, \theta) \) is arbitrary, but this study opens possibilities for expanding models to include asymmetries near the wormhole throat. It is also shown how the scale factor \(a(t) \) relates to such potential asymmetry. Additionally, the influence of the holonomy term decreases with the coordinate \(r\), recovering the FLRW model at a safe distance from the singularity throat.

In summary, this work advances the theoretical understanding of higher-dimensional wormhole models, presenting a comprehensive approach to studying their stability and quantum properties. The findings contribute to the broader understanding of traversable wormholes, with potential implications for future research on semi-integer spin particles and the interplay between geometry, quantum mechanics, and cosmology.

\section{Declarations}



{\bf Authors Contribution:} The author is the sole writer of the manuscript.

{\bf Data Availability Statement:} Data sharing not applicable to this article as no datasets were generated or analyzed during the current study.

{\bf Conflict of Interests:} The author declare no competing interests.

{\bf Ethical Conduct:} The author declare that the research presented in this paper has been conducted with the highest standards of integrity and in accordance with relevant ethical guidelines.








\section{References}



\end{document}